%% file: egads_v2nd-arXiv.tex
\documentclass[12pt]{article}
\usepackage{graphicx}
\usepackage{float}
\usepackage{textcomp}
\usepackage{authblk}
\usepackage{array}    %for use in tables

\newenvironment{Abstract}{\begin{quotation}  }{\end{quotation}}

\def\Acknowledgements{\bigskip  \bigskip \begin{center} \begin{large}
             \bf ACKNOWLEDGEMENTS \end{large}\end{center}}
\input econfmacros.tex
\title{Evaluation of Gadolinium's Action on Water Cherenkov Detector Systems with EGADS}
\input{authorList.tex}

\begin{document}
\begin{titlepage}

\maketitle

\begin{Abstract}
Used for both proton decay searches and neutrino physics, large water Cherenkov (WC) detectors have been very successful tools in particle physics. They are notable for their large masses and charged particle detection capabilities. 
While current WC detectors reconstruct charged particle tracks
over a wide energy range, they cannot efficiently detect neutrons. Gadolinium (Gd) has the largest thermal neutron capture cross section of all stable nuclei and produces an 8 MeV gamma cascade that can be detected with high efficiency. Because of the many new physics opportunities that neutron tagging with a Gd salt dissolved in water would open up, a large-scale R\&D program called EGADS was established to demonstrate this technique's feasibility. EGADS features all the components of a WC detector, chiefly a 200-ton stainless steel water tank furnished with 240 photo-detectors, DAQ, and a water system that removes all impurities from water while keeping Gd in solution. In this paper we discuss the milestones towards demonstrating the feasibility of this novel technique, and the features of EGADS in detail.
\end{Abstract}
\vfill
\vfill
\end{titlepage}
\def\thefootnote{\fnsymbol{footnote}}
\setcounter{footnote}{0}

\section{Introduction}

Water Cherenkov (WC) detectors typically contain a large number of protons, both to study their possible decay as well as to present a large target for neutrino interactions. These detectors are now a well-established tool for conducting particle physics research. However, neutrons cannot be efficiently detected~\cite{Zhang:2013tua} in WC detectors. Free neutrons in these detectors are first thermalised and then mostly captured on protons within about 200 $\mu$s (neutron capture cross section on free protons is 0.3 barns while on oxygen it is 0.19 millibarns). The capture on a proton produces a single 2.2 MeV gamma that is very difficult to detect because the Compton scattered electron is relatively close to Cherenkov threshold, and so produces too few photons given the typical photocathode coverage in WC detectors. In addition, at these low energies there are many background processes present, in particular those produced by radon and spallation.

In 2003, GADZOOKS! was proposed, the idea of enriching WC detectors with a water soluble gadolinium (Gd) salt~\cite{Beacom:2003nk}. Naturally occurring Gd has the largest cross section for the capture of thermal neutrons of all the naturally occurring elements ($\sim$49000 barns). The largest contributions come from the two isotopes $^{157}$Gd and $^{155}$Gd, with about 255,000 and 61,000 barns respectively, and natural abundances of 15.65$\%$ and 14.80$\%$. After neutron capture on $^{157}$Gd and $^{155}$Gd, gamma cascades follow with total energies 7.9 MeV and 8.5 MeV, respectively. Hereafter, we will collectively refer to these gamma cascades as 8 MeV gamma cascades.

Gd is insoluble in water but there are Gd compounds that could be used. Gd nitrate, Gd(NO$_3$)$_3$, has been used as a neutron poison in nuclear reactors but nitrates are mostly opaque in the UVA region~\cite{Jiao:2013} which covers a large portion of the effective spectrum. %http://www.asianjournalofchemistry.co.in/User/ViewFreeArticle.aspx?ArticleID=25_4_116
Gd chloride, GdCl$_3$, is easily soluble and has good Cherenkov light transparency. Gd sulfate, Gd$_2$(SO$_4$)$_3$, has a similar solubility and transparency, and in addition it is less reactive  than GdCl$_3$ and thus more suitable to be used in a detector. Therefore, we chose Gd sulfate. Gd sulfate is easier to dissolve when octahydrated: Gd$_2$(SO$_4$)$_3$ $\cdot$ 8H$_2$O (8 molecules of water per Gd atom). Hereafter, we will refer to it as just Gd sulfate and omit that it is octahydrated.

\begin{figure}[htb]
\centering
\includegraphics[height=2.5in]{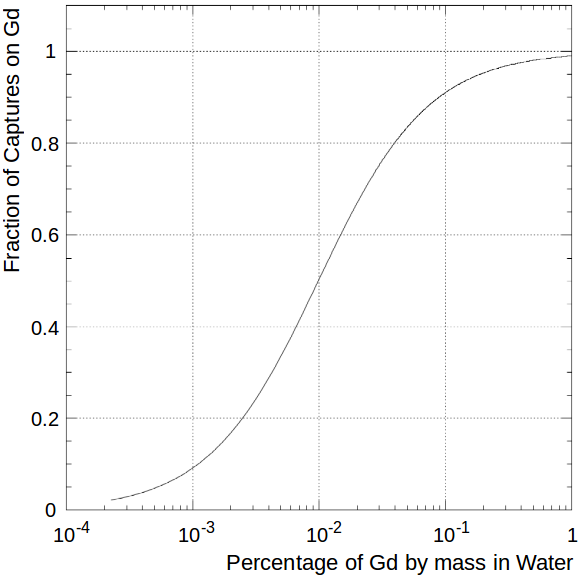}
\caption{Fraction of neutrons captured on Gd as a function of its concentration in water by mass.}
\label{fig:Gdconcentration-vs-captures}
\end{figure}

To achieve 90$\%$ of the neutron captures on Gd after dilution, we need to achieve a concentration of about 0.2$\%$ of Gd sulfate by mass, i.e. about 0.1$\%$ of dissolved Gd; see Figure~\ref{fig:Gdconcentration-vs-captures}. This means we will need to dissolve about 100 tons of Gd sulfate into the 50 kton Super-Kamiokande (Super-K, SK) to achieve this goal. With this Gd concentration, neutrons thermalise and are then captured within about 30 $\mu$s (see Figure~\ref{fig:neutronCaptureTime} in section~\ref{subsec:running_EGADS}).

%%%%%%%%%%%%%%%%%%%%%%%%%%%%%%%%%%%%%%%%%%%%%%%%%%%%%%%%%%%%%%%%%%%%%%%%%
%%%%%%%%%%%%%%%%%%%%%%%%%%%%%%%%%%%%%%%%%%%%%%%%%%%%%%%%%%%%%%%%%%%%%%%%%%

\subsection{Physics}
\label{subsec:physics}

Neutron tagging in WC detectors opens up many new possibilities because it is a powerful tool to reduce backgrounds. Here we will discuss some of the physics topics most likely to benefit from efficient neutron tagging in WC detectors: galactic supernova neutrinos, diffuse supernova neutrino background, atmospheric neutrinos, and proton decay analyses. In addition, though not considered here, Gd loading is also expected to enhance both long-baseline and reactor neutrino studies.

{\bf Galactic supernova neutrinos and diffuse supernova neutrino background}

A core-collapse supernova (SN) releases about 10$^{46}$ J. Out of this vast amount of energy, $\sim$99$\%$ is released in neutrino production. Since neutrinos interact with matter only weakly, they leave the exploding star and travel through space without significant attenuation. Neutrino detectors like Super-K can easily detect galactic supernova explosions (SNe) through inverse beta decay (IBD) events $\bar{\nu}_e + p \rightarrow n + e^+ $. Because of the large cross section and the relatively large positron energies involved, these events represent about 88$\%$ of the total events~\cite{Takahashi:2001ep, Totani:1998nf}. Elastic scattering events, $\nu_x + e^- \rightarrow \nu_x + e^-$, where $\nu_x$ are neutrinos and antineutrinos of all species, represent only about 3$\%$ of all the events~\cite{Takahashi:2001ep}, but are very useful because they point back to the SN. Since we cannot distinguish electrons from positrons (positron annihilation is entirely invisible in WC detectors), directional elastic scattering events are diluted by non-directional IBD events which limits the pointing accuracy. By efficiently identifying IBD events by the observation of neutron captures on Gd and removing them from the elastic scattering sample, the pointing accuracy doubles~\cite{Tomas:2003xn, Beacom:1998fj}. As a consequence, the area of the sky in which astronomers would expect to eventually observe the SN would be reduced by a factor of four.

Among other benefits, a Gd-loaded Super-K would have enhanced sensitivity to late black hole formation~\cite{Beacom:2000qy} and extend the neutrino observation of the cooling phase to later times. Also, efficient neutron detection opens up the possibility to see stellar neutrinos from silicon fusion in nearby massive stars~\cite{Odrzywolek:2003vn, Simpson:2019xwo} (distance less than 1 kpc and M $>$ 13 M$\odot$). This is the last phase in the lifetime of a massive star and lasts from a few hours to a few days before the stellar core collapse, so it serves as a pre-supernova warning.

\begin{figure}[htb]
\centering
\includegraphics[height=3.33in]{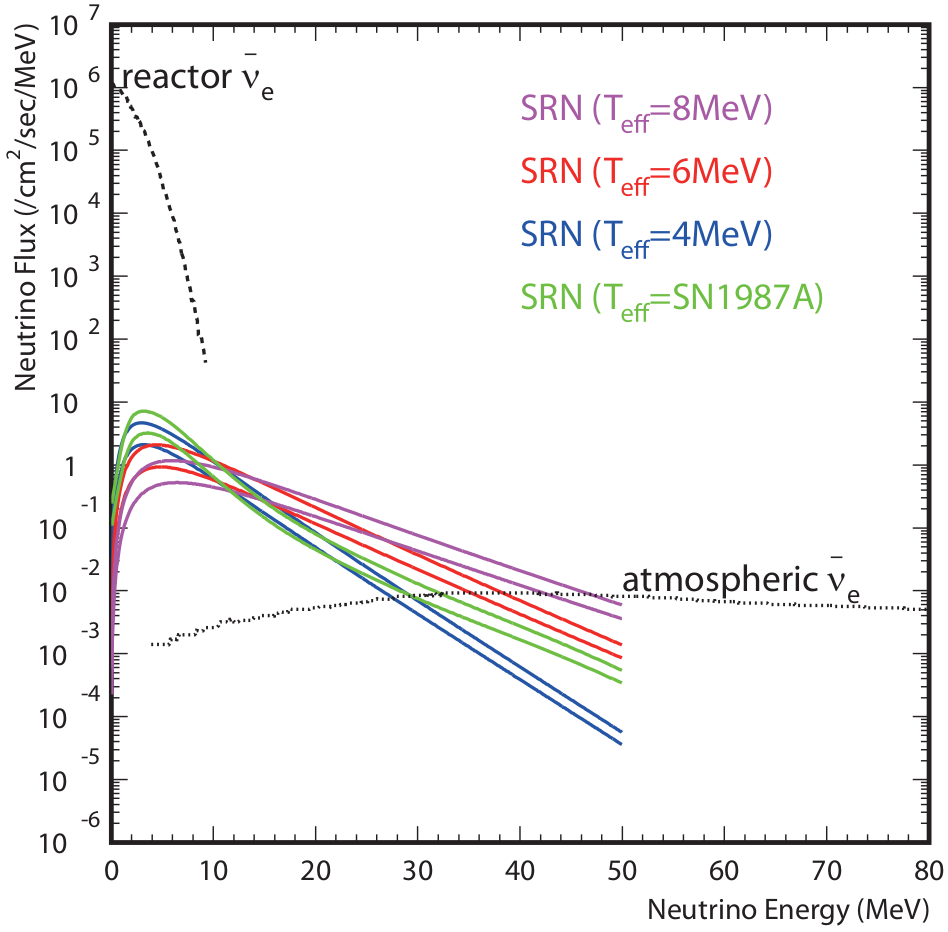}
\caption{Energy spectrum of the diffuse supernova neutrino background for several effective neutrino temperatures~\cite{Horiuchi:2008jz}, T$_{eff}$. The spectra of reactor and atmospheric antineutrinos are also shown.}
\label{fig:DSNB-spectrum}
\end{figure}

We expect about two or three SNe per century in our galaxy~\cite{Adams:2013ana}. The neutrino flux of a single SN far from our galaxy is not large enough to be detected. However, there have been many SNe in the history of the universe, creating a copious, ubiquitous and isotropic neutrino flux: the diffuse SN neutrino background (DSNB)~\cite{Beacom:2003nk}. The predicted spectra are shown in Figure~\ref{fig:DSNB-spectrum}. In this case, we would not be able to link a given event in our detector with a specific SN. However, if we could collect enough events we would acquire information about the neutrino spectrum of an average SN, the history of stellar formation and collapse, the percentage of optically failed SNe, the universe's expansion rate, and would establish the most stringent constrains to neutrino decay.

%%%%%%%%%%%%%%%%%%%%%%%%%%%%%%%%%%%%%%%%%%%%%%%%%%%%%%%%%%%%%%%%%%%%%%%%%
%%%%%%%%%%%%%%%%%%%%%%%%%%%%%%%%%%%%%%%%%%%%%%%%%%%%%%%%%%%%%%%%%%%%%%%%%%

The DSNB has not yet been observed, though the current best limits have been set by Super-K~\cite{Bays:2011si}. Reactor and atmospheric antineutrinos limit the search below 8 MeV and above 30 MeV, respectively, which defines the search window. However, this analysis is limited by currently irreducible backgrounds in this window. These backgrounds would be greatly reduced by requiring the distinctive coincident prompt/delayed signals arising from efficient neutron tagging capabilities. In addition, we would be able to lower the current energy threshold in the analysis. Reactor antineutrinos with energies up to about 8 MeV would impose an upper limit to observing DSNB from SNe with redshifts of about $z=1$. After adding gadolinium sulfate to Super-K, we expect to record up to six DSNB events per year~\cite{Beacom:2003nk}.

{\bf Atmospheric neutrinos}

\begin{figure}[htb]
\centering
\includegraphics[height=1.33in]{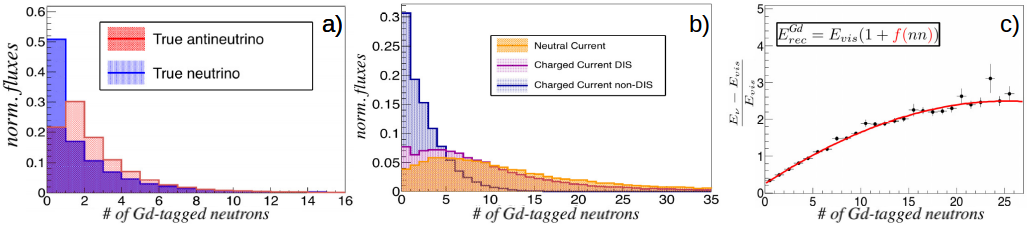}
\caption{Monte Carlo study of a) neutron multiplicity production for neutrinos and antineutrinos; b) NC, CC DIS (Deep Inelesatic Scattering), and CC non-DIS interactions; c) energy correction to the visible energy as a function of the neutron multiplicity.}
\label{fig:AtmNeutr_neutron_tagging}
\end{figure}

Efficient neutron tagging adds more information about the neutrino, the interaction type and the neutrino energy. To give an idea of the potential information gain from efficient neutron tagging, a Monte Carlo (MC) study was done. The MC simulates 500 years of atmospheric neutrinos at Super-K where the flux was taken from~\cite{Honda:2015fha} and neutrino primary interactions were calculated with NEUT~\cite{Hayato:2009zz}. The gamma cascade spectrum from neutron captures on Gd was calculated using GEANT4~\cite{Agostinelli:2002hh} while secondary interactions were simulated with GEANT3~\cite{Zeitnitz:1994bs}. Three examples of these simulations are shown in Figure~\ref{fig:AtmNeutr_neutron_tagging}:

- Neutrino-antineutrino separation: although the neutron multiplicity increases with energy due to interactions inside the nucleus, the neutron multiplicity is larger for antineutrinos than for neutrinos. See Figure~\ref{fig:AtmNeutr_neutron_tagging}a.

- Neutral and charged current separation: neutral current (NC) interactions are non-flavor changing interactions and thus carry no flavor information while charged current (CC) interactions do. In oscillation analyses, NC is an important contamination source of e-like events. NC events tend to deposit a larger amount of energy in the target nucleus than CC interactions. As a consequence, they produce on average larger numbers of neutrons as shown in Figure~\ref{fig:AtmNeutr_neutron_tagging}b.

- Neutron-corrected neutrino energy: accurate neutrino energy reconstruction is very important in the atmospheric oscillation analysis. Figure~\ref{fig:AtmNeutr_neutron_tagging}c shows the fraction of non-visible energy as a function of the number of tagged neutrons on simulated atmospheric neutrino interactions at Super-K, where E$_{\nu}$ is the neutrino energy and E$_{vis}$ is the reconstructed energy from charged particles. These neutrinos often interact with the nuclear media, enhancing neutron production; the current analysis cannot take into account the energy consumed in this process. With efficient neutron tagging this feature can be used to reconstruct the neutrino energy, E$_{rec}^{Gd}$, from the visible energy.
%As shown in Figure~\ref{fig:AtmNeutr_neutron_tagging}c, with efficient neutron tagging this feature can be used to recover the neutrino energy, E$_{rec}^{Gd}$, from the visible energy, E$_{vis}$ (reconstructed from charged particles).

%%%%%%%%%%%%%%%%%%%%%%%%%%%%%%%%%%%%%%%%%%%%%%%%%%%%%%%%%%%%%%%%%%%%%%%%%
%%%%%%%%%%%%%%%%%%%%%%%%%%%%%%%%%%%%%%%%%%%%%%%%%%%%%%%%%%%%%%%%%%%%%%%%%%

{\bf Proton decay}

Large WC detectors were first built in the early 1980's with proton decay searches inspired by SU(5)-based predictions~\cite{Pati:1973rp, Georgi:1974sy} as their primary motivation. The main background for proton decay analyses comes from atmospheric neutrino interactions. As shown in Figure~\ref{fig:AtmNeutr_neutron_tagging}a and \ref{fig:AtmNeutr_neutron_tagging}b, atmospheric neutrinos often produce at least one neutron. On the other hand, and taking the canonical mode $p \rightarrow e^+ \pi^0$ as a reference, few neutrons are expected to be produced in a proton decay. %: only 7$\%$ of the events are expected to be accompanied by one neutron or more. 
A significant sensitivity gain for the proton lifetime limit determination when using neutron tagging is therefore expected.

%Mesmer \index{Mesmer}
%Table~\ref{tab:blood}.

\subsection{EGADS demonstrator}
\label{subsec:egadsdemonstrator}

GADZOOKS! ({\bf G}adolinium {\bf A}ntineutrino {\bf D}etector {\bf Z}ealously {\bf O}utperforming {\bf Old} {\bf K}amiokande, {\bf S}uper{\bf !}) was first proposed in late 2002, and soon after investigations into its application at Super-K began at the University of California, Irvine (UCI), and in particular on how to purify Gd-loaded water. UCI also developed a device to precisely monitor the water transparency resulting from that water purification system.

\begin{figure}[htb]
\centering
\includegraphics[height=4.0in]{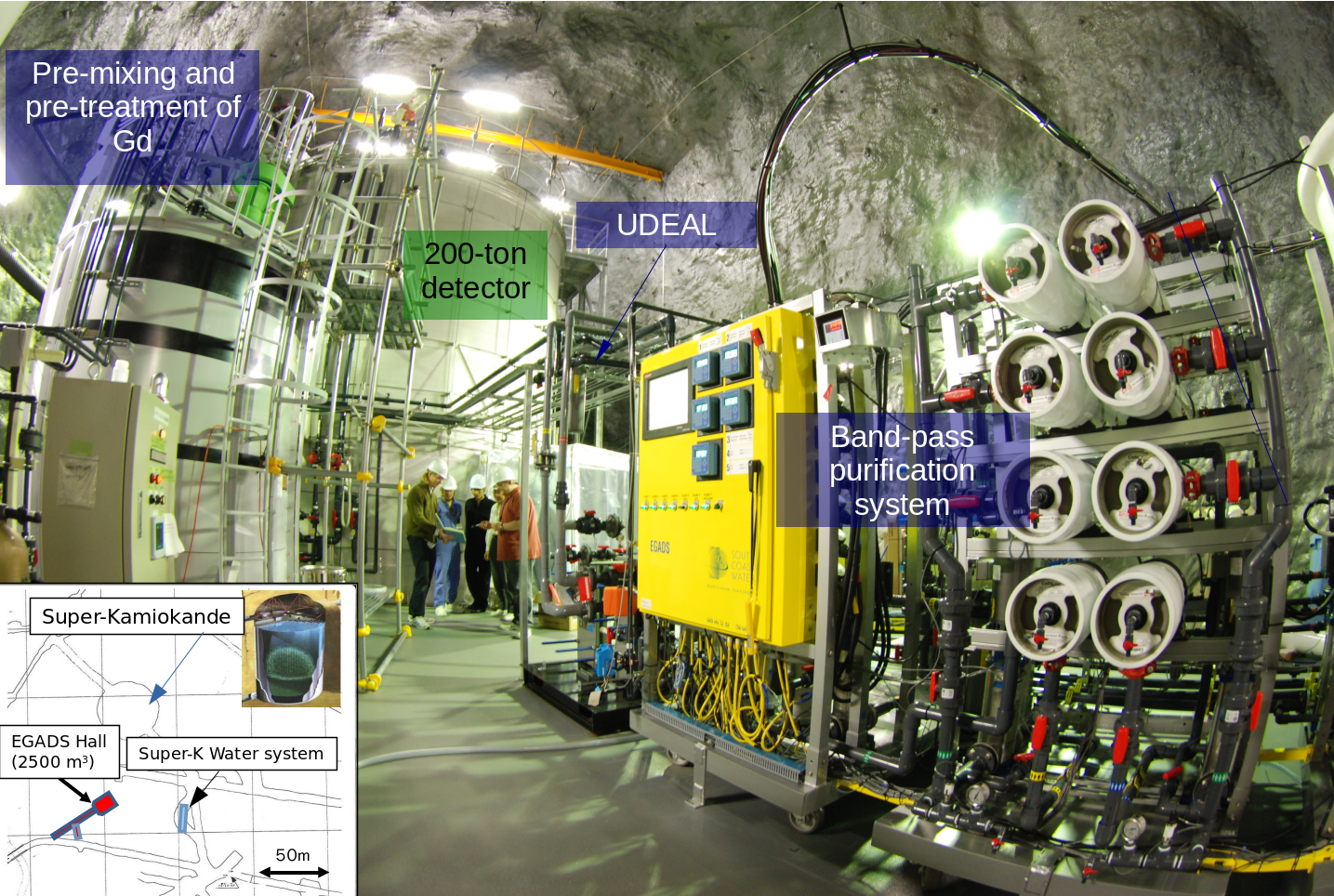}
\caption{Located near Super-K, EGADS initially featured: a 200-ton tank, a pre-treatment system (Gd sulfate dissolving system and its pre-treatment), UDEAL (a transparency measurement device) and the band-pass water purification system.}
\label{fig:EGADS-hall}
\end{figure}

\begin{figure}[htb]
\centering
\includegraphics[height=3.2in]{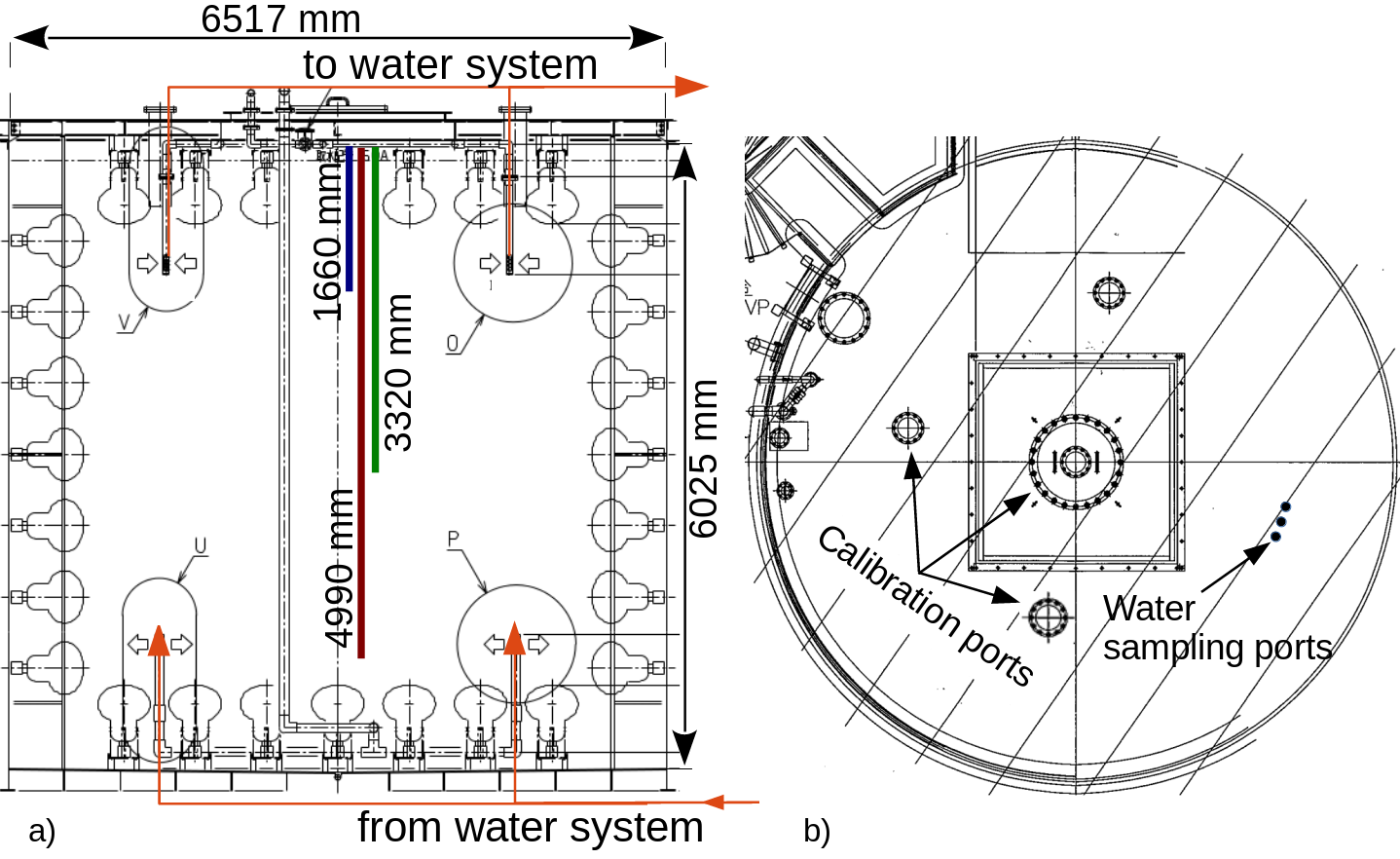}
\caption{Left a) schematic view of the EGADS tank, its sampling ports (bottom, centre and top at 4990, 3320, and 1660 mm from the tank-top) and inflow and outflow from and to the water purification systems. Right b) top view of the EGADS tank with its calibration and water sampling ports.}
\label{fig:EGADS-tank-views}
\end{figure}

An ~important ~milestone ~was ~reached ~when ~in ~2009 ~the ~EGADS ({\bf E}valuating \\
{\bf G}adolinium's {\bf A}ction on {\bf D}etector {\bf S}ystems) project was funded to test the results from the UCI R$\&$D on a larger scale. Located in a newly excavated cavern near Super-K (about 1000 meters underground in the Kamioka mine, Hida-city, Gifu, Japan; see Figure~\ref{fig:EGADS-hall}), EGADS initially featured: a 200-ton stainless steel tank, a pre-treatment system (Gd sulfate dissolving system and its pre-treatment), a water purification system (see Section~\ref{subsec:water-filtration}) and a transparency measurement device called UDEAL ({\bf U}nderground {\bf D}evice {\bf E}valuating {\bf A}ttenuation {\bf L}ength, see Section~\ref{subsec:udeal}). Like Super-K, EGADS has a cylindrical shape (see Figure~\ref{fig:EGADS-tank-views}) but has an inner detector only. Calibration ports are also available at the top. To measure Gd sulfate concentrations, a Hitachi atomic absorption spectrometer (AAS) of the ZA3000 series was used.

%%%%%%%%%%%%%%%%%%%%%%%%%%%%%%%%%%%%%%%%%%%%%%%%%%%%%%%%%%%%%%%%%%%%%%%%%
%%%%%%%%%%%%%%%%%%%%%%%%%%%%%%%%%%%%%%%%%%%%%%%%%%%%%%%%%%%%%%%%%%%%%%%%%%

The EGADS project started with five goals:
\begin{itemize}
\item Demonstrate that the purification system can achieve and maintain good water quality while keeping the Gd concentration constant.
\item Show that Gd sulfate has no adverse effects on the Super-K detector components.
\item Demonstrate that the addition of Gd sulfate will not negatively affect existing Super-K analyses.
\item Study how to reduce the now visible neutron background from spallation, U/Th fission chains in Gd sulfate impurities, ambient neutrons, etc.
\item Prove that Gd can be added/removed in an efficient and economical way.
\end{itemize}

In this paper, we will show EGADS has achieved these goals.

%%%%%%%%%%%%%%%%%%%%%%%%%%%%%%%%%%%%%%%%%%%%%%%%%%%%%%%%%%%%%%%%%%%%%%%%%
%%%%%%%%%%%%%%%%%%%%%%%%%%%%%%%%%%%%%%%%%%%%%%%%%%%%%%%%%%%%%%%%%%%%%%%%%%

\section{EGADS water purification systems}
\label{subsec:water-filtration}

The Super-K water purification system produces ultrapure water with a resistivity very close to 18.2 M$\Omega$-cm, the theoretical maximum. This very high purity is achieved by using several stages including $\mu$-filters, ultrafilters (UF), UV lamps, reverse osmosis (RO), vacuum and membrane degasifiers, and resins that remove impurities, dissolved ions, and kill and remove bacteria~\cite{Ikeda:1982uf, BeckerSzendy:1992hr, Fukuda:2002uc}. This means that if we were to simply add Gd sulfate to Super-K as it now stands, the current purification system would naturally remove the Gd along with all the other impurities.

In order to maintain good water quality it is necessary to have a water purification system that removes all impurities, ions included, {\em except} Gd$^{3+}$ and its anionic partner (Cl$^-$, SO${_{4}}{^{2-}}$, etc.), and thus maintain good water transparency. In EGADS, as in Super-K, water from the top of the detector is circulated through the water purification systems, cleaned, and then injected again into the detector from the bottom (see Figure~\ref{fig:EGADS-tank-views}a). In EGADS there are three different Gd-tolerant water purification systems: the Gd sulfate mixing and pre-treatment, the fast recirculation system, and the band-pass system. These systems usually undergo an annual maintenance to ensure optimal efficiency removing impurities. They are described in the next three subsections.

%A new water purification system had to be developed in order to remove all impurities, ions included, {\em except} Gd$^{3+}$ and its anionic partner (Cl$^-$, SO${_{4}}{^{2-}}$, etc.), and thus maintain good water transparency.

\subsection{Pre-treatment system}
\label{subsec:pre-treatment}

\begin{figure}[htb]
\centering
\includegraphics[height=1.6in]{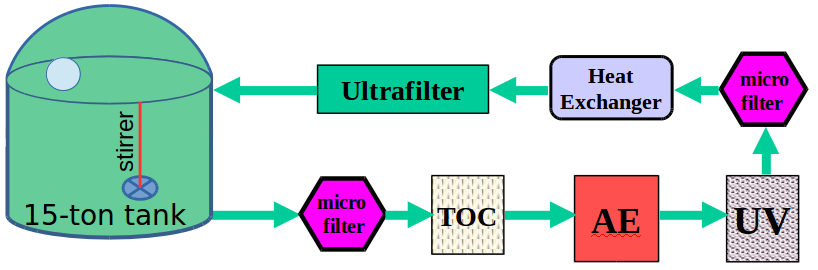}
\caption{Pre-treatment schematic view: a 15-ton tank with a stirrer to dissolve the Gd sulfate, 3 and 0.2 $\mu$m filters, TOC and UV lamps, the AE resin, a heat exchanger, and an ultrafilter.}
\label{fig:EGADS-pre-treatment}
\end{figure}

The pre-treatment system consists of a 15-ton polyethylene mixing tank (including a stirrer to dissolve the Gd sulfate), TOC (Total Organic Compounds) and UV lamps to kill and remove bacteria together with $\mu$-filters, a special strongly basic anion exchange (AE) resin, a heat exchanger, and an UF. This resin was designed and shown to remove uranium with an efficiency above 99$\%$ and with no Gd loss. A schematic view is shown in Figure~\ref{fig:EGADS-pre-treatment}.

%%%%%%%%%%%%%%%%%%%%%%%%%%%%%%%%%%%%%%%%%%%%%%%%%%%%%%%%%%%%%%%%%%%%%%%%%
%%%%%%%%%%%%%%%%%%%%%%%%%%%%%%%%%%%%%%%%%%%%%%%%%%%%%%%%%%%%%%%%%%%%%%%%%%

After filling the 15-ton mixing tank with pure water derived from the XMASS experiment's water purification system~\cite{ABE201378}, Gd sulfate is added. The built-in stirrer then runs until dissolving is complete. Due to the limited size of the mixing tank,  concentrations higher than 0.2$\%$ are often prepared.  Since  the band-pass system is optimized to run at concentrations up to 0.2$\%$, concentrated Gd-loaded water coming from the pre-treatment system has to be diluted -- typically with water from the 200-ton main tank -- back down to this concentration. Then, after going through the band-pass system, Gd-loaded water is injected into the 200-ton tank from the bottom.

\subsection{Band-pass filtration system}
\label{subsec:band-pass}

\begin{figure}[htb]
\centering
\includegraphics[height=3.5in]{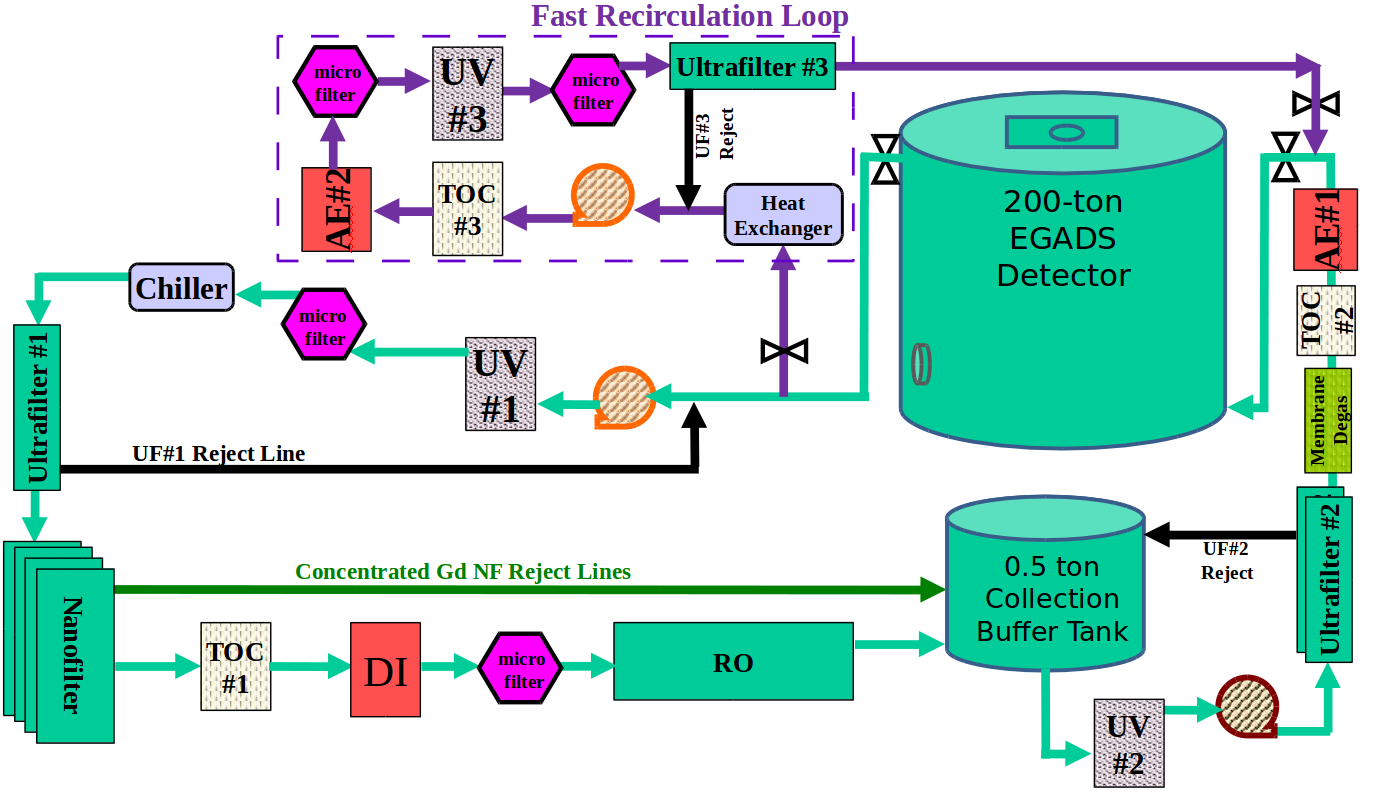}
\caption{Schematic view of the band-pass system and the fast recirculation system (inside dashed line). These systems were built in cooperation with the South Coast Water company.}
\label{fig:filtration_systems}
\end{figure}

The band-pass filtration system (see Figure~\ref{fig:filtration_systems}) is EGADS' main filtration system and under usual conditions it runs in parallel with the fast recirculation system (see Subsection~\ref{subsec:fast_recirculation}). The total flow through both systems is about 90 liters/minute, approximately evenly split between the two.

%%%%%%%%%%%%%%%%%%%%%%%%%%%%%%%%%%%%%%%%%%%%%%%%%%%%%%%%%%%%%%%%%%%%%%%%%
%%%%%%%%%%%%%%%%%%%%%%%%%%%%%%%%%%%%%%%%%%%%%%%%%%%%%%%%%%%%%%%%%%%%%%%%%%

The band-pass system takes water from the top of the 200-ton EGADS detector. The first important elements are a UV unit and $\mu$-filters to kill and trap bacteria before they can enter the system. The second element is a chiller: Gd sulfate dissolves more readily at lower temperatures since this process is exothermic. Later elements (pumps, filters) will increase the water temperature. Thus, to avoid precipitation in key elements, a chiller is needed. The third element is an UF to further remove relatively large impurities.

The fourth element is a series of nanofilters (NF). With the proper pass/reject flow settings these can be used to split the flow between Gd-less water (well below 1 ppm) and concentrated Gd water from the NF reject. This last line directly goes into a 0.5 ton collection buffer tank. The other line contains basically no Gd and hence we proceed to remove all remaining impurities. If any Gd is present in this line, it will be removed from the system along with all the other impurities through deionization (DI) and RO. This produces ultrapure water; about 15 liters/minute undergo this treatment. At the end of it, this line joins the concentrated Gd water from the NF reject line in the 0.5 ton collection buffer tank. 

From the 0.5 ton collection buffer tank, in addition to similar elements described above, there is a membrane degasifier (to remove dissolved air), a TOC lamp and finally the AE resin before water is injected back into the bottom of the EGADS detector. The AE resin was initially used to remove uranium (see Subsection~\ref{subsec:pre-treatment}). In addition to this, it was discovered that this resin also improves the water transparency LL15 value by about 3$\%$ (the LL15 metric is defined in Section~\ref{subsec:udeal}). %pg23 May18,2012 CM Vagins
The flow inside the EGADS detector is from the bottom (purified water) to the top; it takes about two days to turn over the entire water volume.

We also found that when using new membranes, it is crucial to thoroughly flush them in advance. For this task we built a dedicated flushing system with a large buffer tank and DI resin unit. The goal is to completely remove the preservative liquids that are used when these membranes are packaged at their factories, thereby achieving proper conditioning of the membranes before usage in our water filtration systems.

\subsection{Fast recirculation system}
\label{subsec:fast_recirculation}

The fast recirculation system (see Figure~\ref{fig:filtration_systems}) provides increased cleaning power, albeit not as powerful as the band-pass system since it has no element to remove cations. It consists of most of those elements that cannot remove Gd: TOC, UV, AE resin, and UF. In addition, it is equipped with a heat exchanger to remove the heat from these elements. Note that lower temperatures allow easier control of bacterial growth as well as a higher solubility of Gd sulfate.

\section{Monitoring water transparency}
\label{subsec:udeal}

Water quality is a key parameter of any WC detector. It ensures good water transparency and, as a consequence, that the Cherenkov light attenuation length is long compared to the detector size. In such a case, even charged particles at low energies creating few photons can be efficiently detected. In addition, while usually present at levels much too small to directly affect water transparency, trace amounts of radioactive impurities in the water can increase backgrounds via decays that mimic true physics signals. Removing such impurities is vital since they can easily make events within the detector's fiducial volume.

The more abundant kinds of impurities (dissolved iron, amines, bacteria, etc) typically reduce the water transparency by increasing light absorption and/or scattering, and this has to be monitored regularly. Thus, a dedicated device was developed to measure water transparency at UCI. UDEAL ({\bf U}nderground {\bf D}evice {\bf E}valuating {\bf A}ttenuation {\bf L}ength) was built based on this device.

\begin{figure}[htb]
\centering
\includegraphics[height=3.in]{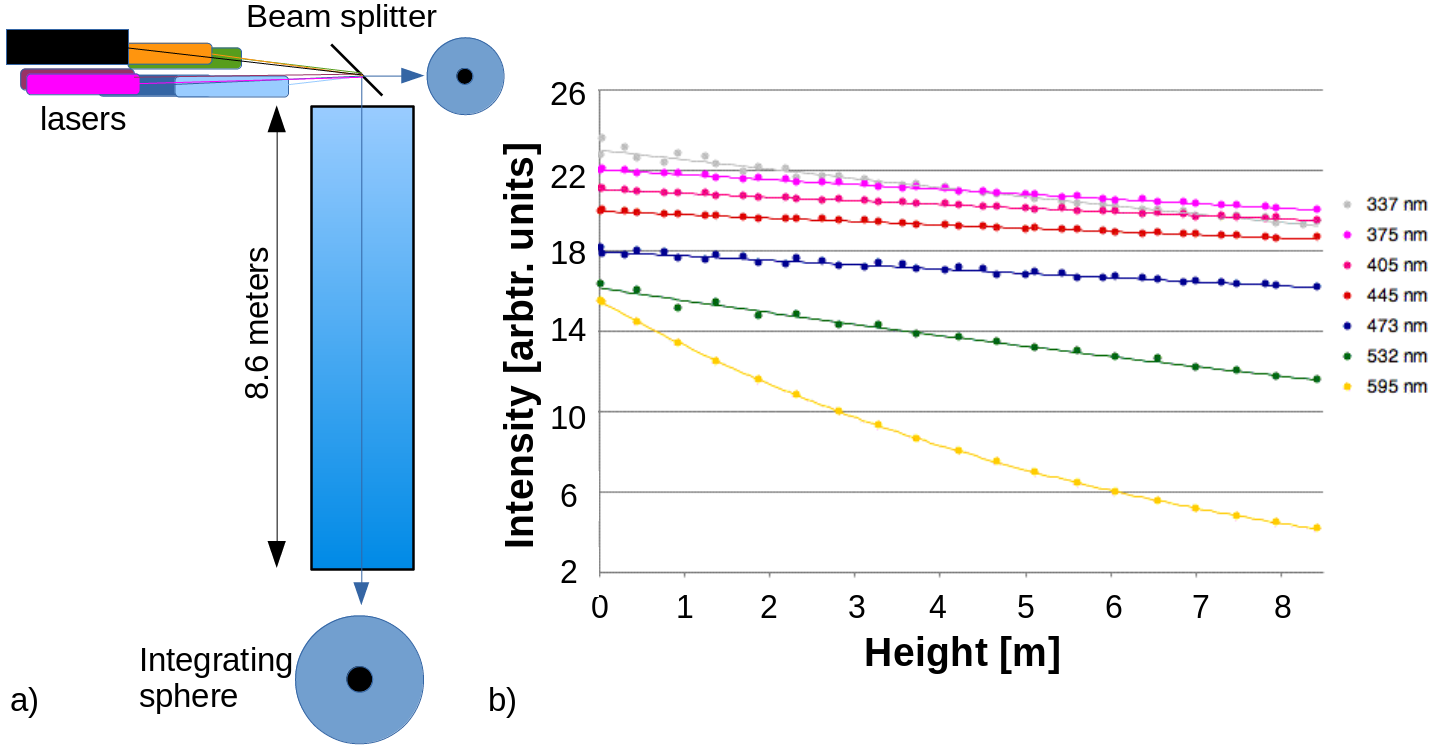}
\caption{On the left a) schematic view the water transparency monitor device, UDEAL: beam injector and splitter with an integrating sphere to monitor the beam intensity (top), a pipe (water height is adjustable) and integrating sphere to measure the beam intensity at the bottom. Right b), example of water transparency measurement: light intensity for the seven beam wavelengths as a function of water height.}
\label{fig:UDEAL}
\end{figure}

UDEAL consists of three basic components as shown in Figure~\ref{fig:UDEAL}a: a light injector for seven laser beams and intensity monitor (integrating sphere, top right), a pipe (8.6 meters tall, middle) and an integrating sphere to monitor the transmitted light (below the pipe). The beam injector consists of seven lasers (3 laser pointers, 3 professional laser diodes and a nitrogen laser) spanning a range of wavelengths matched to the Cherenkov spectrum and phototube response function: 337, 375, 405 445, 473, 532, and 595 nm. UDEAL operates only one of these laser beams at a time, and automatically cycles through all seven in a pre-determined sequence. To reduce beam jitter effects on the light intensity measurements, integrating spheres randomize the position and direction of the detected laser light. These custom-build integrating spheres (the one at the top is a 10-cm diameter sphere while the one at the bottom is 30 cm) are coated on the inside with a special diffuse high reflectivity paint. They are read out by 10x10 mm UV-enhanced silicon photodiodes inlaid in the sphere walls. A baffle shields the photodiodes from seeing the beam spot on the inside of the sphere walls, which ensures that there are at least two diffuse reflections before a photon can contribute to a signal.

%%%%%%%%%%%%%%%%%%%%%%%%%%%%%%%%%%%%%%%%%%%%%%%%%%%%%%%%%%%%%%%%%%%%%%%%%
%%%%%%%%%%%%%%%%%%%%%%%%%%%%%%%%%%%%%%%%%%%%%%%%%%%%%%%%%%%%%%%%%%%%%%%%%%

The beam (top left) hits a beam splitter where part of it is directed to the pipe (downwards) and part of it is directed to an integrating sphere (top right). This first integrating sphere monitors the beam intensity fluctuations. The pipe is automatically filled with water to different heights. For each height, measurements of the beam intensities at the bottom integrating sphere are performed. Hence, there is a measurement of the intensity difference for each height and wavelength (see Figure~\ref{fig:UDEAL}b). These measurements are conducted both while filling and draining the pipe to reduce the impact of several time dependent systematic effects. While the measurements are taken, the water height is kept constant. With these measurements we can then calculate the absolute light attenuation length in water for each wavelength.

To monitor the water quality, we compared the changes in water transparency for every wavelength for pure water and water with different Gd sulfate concentrations, as well as with different filtration systems and setups. To make these comparisons easier we often combine the seven wavelength measurements into a single number. In order to define this number, we used the Geant3 based Super-K MC to simulate electrons uniformly distributed throughout the detector volume with random directions. The Cherenkov photons were then tracked and the travel distances recorded as well as the wavelengths of the recorded photons by every photo-multiplier (PMT). The average photon travel distance in Super-K was determined to be 15.2 meters. Therefore, the percentage of Cherenkov light left after a travel distance of 15 meters (LL15) - what would remain available to activate SK's phototubes after passing through the fluid in question - was used as a standard comparison for different Gd sulfate loading concentrations, different filtration setups or any other change in experimental conditions. For a given wavelength, the light left at 15 m can be calculated from the attenuation length and a simple exponential.

To determine the Cherenkov LL15, the wavelengths of the recorded photons by every PMT from the MC were used. The relevant wavelengths are between 300 and 600 nm, % and their distribution was binned such that the bin boundaries are setas the average of each of the neighboring two laser wavelengths from UDEAL. 
the same wavelengths covered by the UDEAL lasers. With this information, we can determine the average fraction of Cherenkov light which is represented by each of the wavelengths used in UDEAL. The fractions were determined to be: 0.25, 0.25, 0.21, 0.14, 0.11, 0.04, and 0.003 for the UDEAL wavelengths, 337, 375, 405, 445, 473, 532, and 595 nm, respectively. A weighted sum of all the LL15 for each wavelength is then calculated. This is the Cherenkov light left after travelling 15 m with PMT efficiencies included: Cherenkov LL15 = $\sum_{\lambda_i} W_i ~\mathrm{e}^{-15/L_i}$, where the sum runs over the seven wavelengths ($\lambda_i$) and $W_i$ and $L_i$ are the fractions of Cherenkov light and attenuation lengths for a given wavelength, $\lambda_i$.

%For the water in Super-K, LL15 has been calculated since 2006. It has varied within 74.7$\%$ and 82.1$\%$, i.e. between about 75$\%$ and 82$\%$ of the Cherenkov photons survive their passage through SK's ultrapure water to reach the inner detector (ID) PMTs and potentially generate photoelectrons.

% Rayleigh omitted

UDEAL was designed to take data automatically every day. Since the water purification system supplies clean water at the bottom of the tank and draws from the top, a water transparency gradient is expected: cleaner at the bottom and more impurities as we move to the top. Hence, these measurements are done by taking samples from three different ports inside the tank situated at the top, center, and bottom of the tank, 4990, 3320, and 1660 mm from the tank top respectively, see Figure~\ref{fig:EGADS-tank-views}a. One LL15 measurement per day per sampling position covering all wavelengths is taken.

\section{Potential impact on Super-K detector components}

The 200-ton EGADS main tank is made of the same stainless steel as the 50-kton Super-K tank, SUS 304. All the other components that were later installed and that are in contact with water are also made with the same materials as were used in Super-K. Examples are PMTs, acrylic covers and fiber-glass reinforced plastic (FRP) cases to avoid a PMT implosion chain reaction, cables, shrink tubes for PMT connector protection, screws, etc. In this section, results of soak tests are presented. 

\subsection{Soak tests for current Super-K materials}

Soak tests of all materials used at Super-K were done in the USA and Okayama University in Japan. The procedure for the soak tests was the following:

\begin{itemize}
\item Soak each material both in Gd-loaded water and Super-K pure water at room temperature (around 25\textdegree C, while Super-K temperature water is about 13\textdegree C).
\item With an spectrophotometer (JASCO V-550 with wavelengths between 190 and 900 nm and 2 cm cells), measure the transparency of the solution (typically after about 3 months of soaking).
\item Estimate the impact of each material in the attenuation length at Super-K (taking into account temperature and volume/surface ratios of the samples as compared to those in Super-K).
\end{itemize}

The Super-K purification system can achieve and maintain ultrapure water quality as has been demonstrated after running for many years. The materials were soaked in both Gd-loaded water and pure Super-K water: if a given material steadily emanates a measurable amount of impurities in pure water, it has been established that the Super-K water purification system can keep up with removing them. Similarly, any potential adverse effect of the material being soaked in Gd sulfate-loaded water will be reduced via continuous circulation and filtration.

Due to the variability and availability of materials and differences among manufacturers, soak tests of all materials to be used in any future WC detector should be done in ultrapure and Gd-loaded water.

\subsection{Soak tests results}

Among all the materials only two showed a measurable effect after the soak test: the Super-K inner detector (ID) cables and the black rubber friction pads used to hold the ID PMTs; the rubber showed the largest effect. Although the effect is stronger with Gd sulfate, the effect is clearly seen in pure water as well but has been unnoticed until now after more than 20 years of data taking with Super-K. This means that the current SK water purification system can take care of the impurities coming from this material through continuous water circulation and purification. Taking into account the surface/volume ratio difference between the soak tests and the actual ratio at Super-K and other effects, it was concluded that even this case does not represent a potential problem. In addition, the high water quality in EGADS, in which the black rubber is also present (as will be discussed in Subsection~\ref{subsec:EGADS-transparency}), demonstrates this point.

\section{Running EGADS}
\label{subsec:running_EGADS}

As mentioned in Section~\ref{subsec:egadsdemonstrator}, among the goals of the EGADS project was to demonstrate that the purification system can achieve and maintain good water quality. We proceeded systematically. 

We started with pure water in the uninstrumented EGADS 200-ton stainless steel tank in the first half of 2011. The band-pass system was able to achieve the goal with ultrapure water. 

Next, from mid-2011 to the end of 2012 we loaded 0.2$\%$ Gd sulfate in the 15-ton mixing tank, which is made of plastic. The band-pass system filtering this Gd-loaded water was also able to achieve and maintain good water quality.

In 2013 we loaded the still uninstrumented 200-ton tank with 0.2$\%$ Gd sulfate with positive results again.

\begin{figure}[htb]
\centering
\includegraphics[height=3.in]{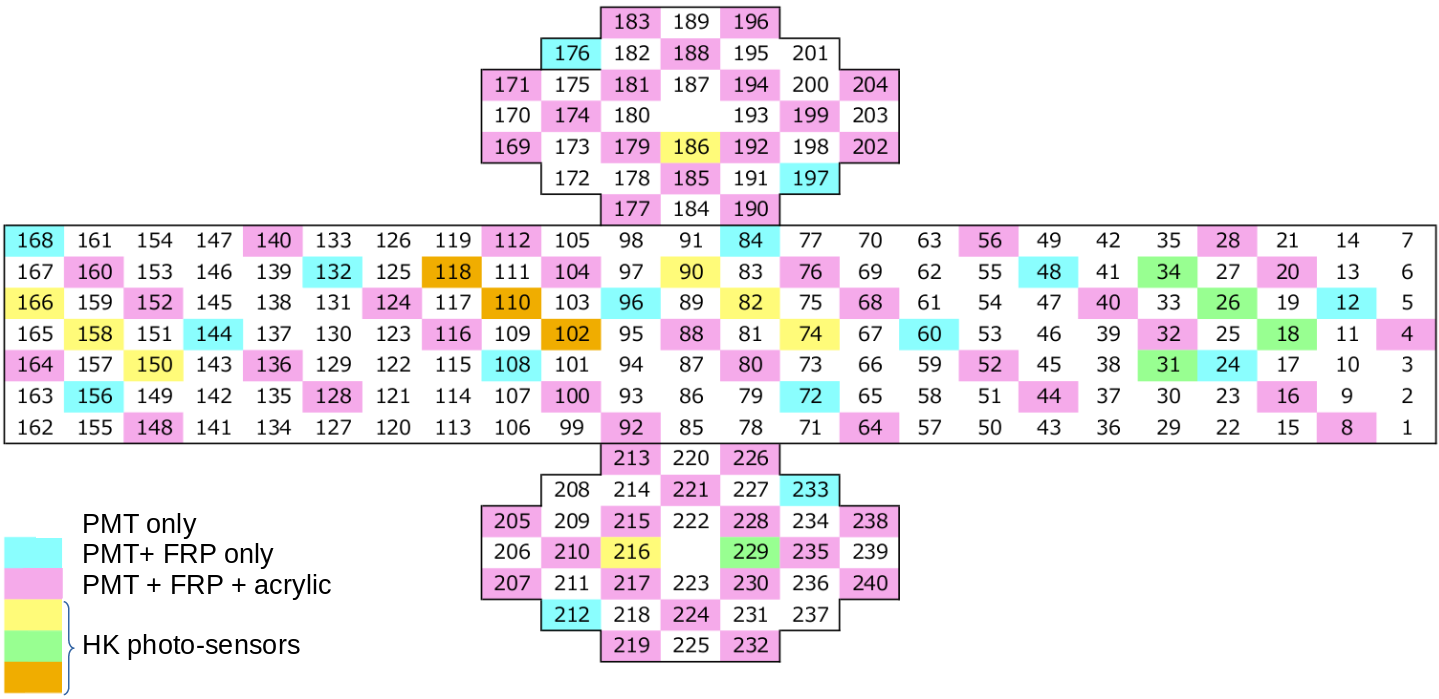} %PMT_ID_expand}
%\vspace{-4.cm}
\caption{EGADS PMT types and covers map. This is the unrolled, inner view of the cylindrical detector.}
\label{fig:PMT_map}
\end{figure}

After the previous successes, in the summer of 2013 we decided to install 240 PMTs in the EGADS tank. Out of these 240, 224 are the 50-cm Super-K ID PMTs~\cite{Suzuki:1992as}, while 16 are several types of photosensors for a Hyper-Kamiokande (HK) R$\&$D project~\cite{Abe:2018uyc, Nishimura:2013cua}. Among the 224 50-cm PMTs: 148 PMTs have no cover, 16 PMTs have an FRP housing only, and 60 PMTs have both an FRP and acrylic cover (same as all ID PMTs at Super-K). All PMT types, covers, and relative positions are shown in Figure~\ref{fig:PMT_map}. Similar to Super-K's ID, the active photocathode coverage is about 40$\%$. As in the Super-K ID, the remainder is covered by black polyethylene terephthalate sheets. There is no outer detector (OD) in EGADS.

%%%%%%%%%%%%%%%%%%%%%%%%%%%%%%%%%%%%%%%%%%%%%%%%%%%%%%%%%%%%%%%%%%%%%%%%%
%%%%%%%%%%%%%%%%%%%%%%%%%%%%%%%%%%%%%%%%%%%%%%%%%%%%%%%%%%%%%%%%%%%%%%%%%%

A magnetic field compensation coil was also activated, reducing the terrestrial magnetic field to less than 0.1 Gauss inside the EGADS tank. The front-end and read-out electronics and related software, as well as the trigger system that was installed in EGADS during 2013 were reused Super-K-I/II/III ATMs~\cite{Tanimori:1988qi}. The high voltage system is a CAEN SY1527LC and an AP1932.

\begin{figure}[htb]
\centering
\includegraphics[height=2.in]{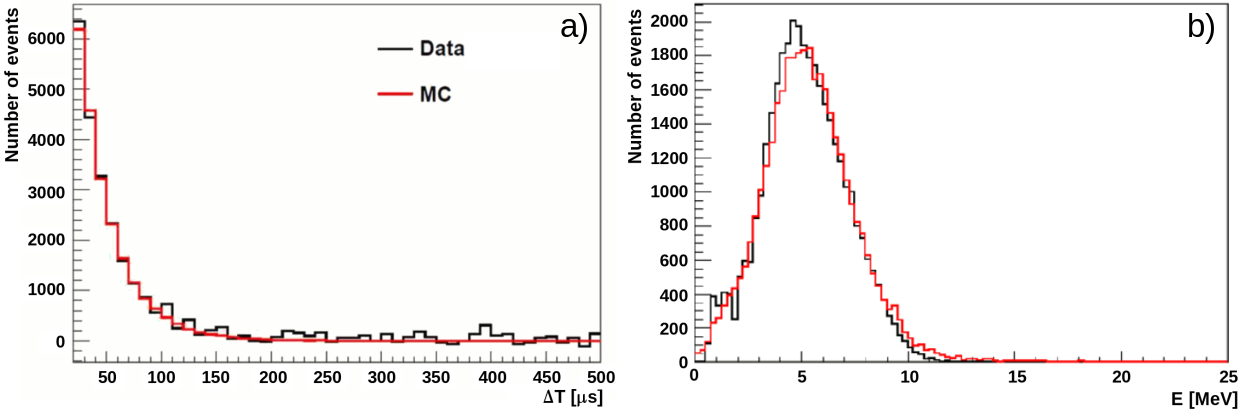}
%\vspace{-4.cm}
\caption{Left a) delayed neutron capture with an Am/Be source for a Gd sulfate concentration of 0.2 $\%$. Right b) reconstructed energy from the delayed neutron signals.}
\label{fig:neutronCaptureTime}
\end{figure}

The installation of PMTs and their ancillary electronics turned the EGADS 200-ton tank into a proper detector. As an example, here we show data taken with an Am/Be source at EGADS in Figure~\ref{fig:neutronCaptureTime}. Figure~\ref{fig:neutronCaptureTime}a shows the time difference between the prompt event and the delayed neutron capture with an Am/Be source for a Gd sulfate concentration of 0.2 $\%$. The fitted neutron capture time here for the data is 29.9$\pm$0.3 $\mu s$ and for MC is 30.0$\pm$0.8 $\mu s$. Figure~\ref{fig:neutronCaptureTime}b shows the reconstructed energy from the delayed neutron signals.

% http://iopscience.iop.org/article/10.1088/1742-6596/718/6/062070/pdf

%%%%%%%%%%%%%%%%%%%%%%%%%%%%%%%%%%%%%%%%%%%%%%%%%%%%%%%%%%%%%%%%%%%%%%%%%
%%%%%%%%%%%%%%%%%%%%%%%%%%%%%%%%%%%%%%%%%%%%%%%%%%%%%%%%%%%%%%%%%%%%%%%%%%

\subsection{Transparency and Gd concentration}
\label{subsec:EGADS-transparency}

\begin{figure}[htb]
\centering
\includegraphics[height=3.5in]{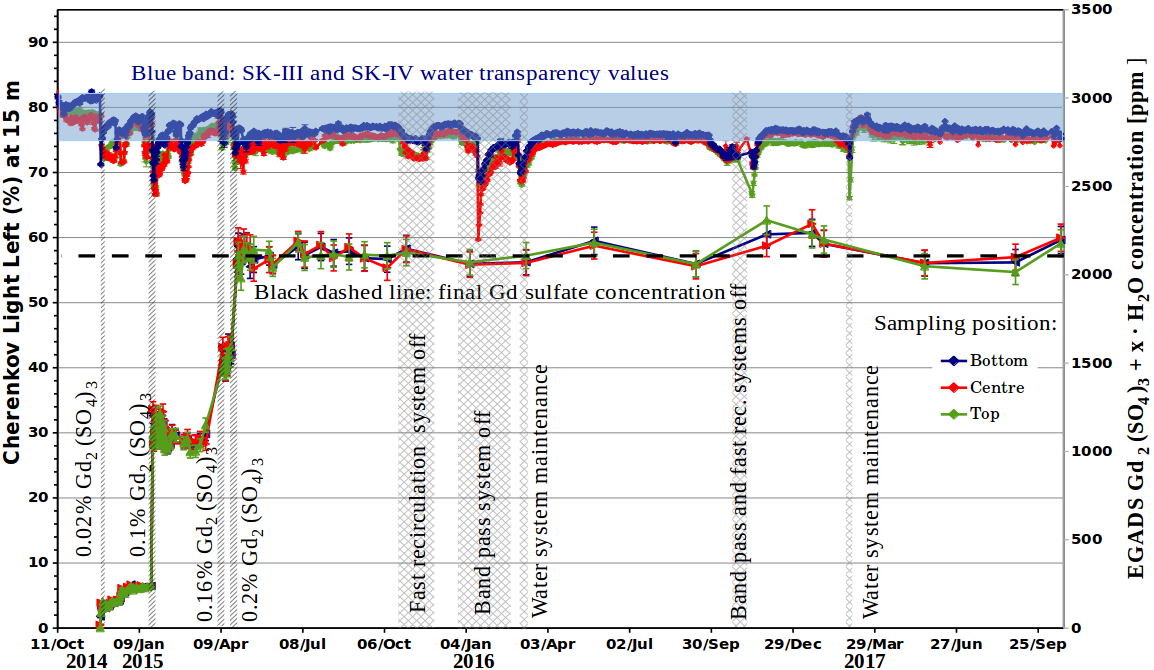}
%\vspace{-4.cm}
\caption{Upper three lines: Cherenkov light left [$\%$] at 15 m (left y-axis scale),  and lower three lines: Gd sulfate concentration for the three sampling positions in the EGADS detector (right y-axis scale). The line colors for the bottom, center and top sampling positions are blue, red and green, respectively. The blue band represents the typical Super-K ultrapure water LL15 values while the horizontal dashed line represents the Gd sulfate final target concentration. }
\label{fig:Transparency-Concentration}
\end{figure}

In 2014 we started loading Gd into the EGADS detector. Because at this point EGADS mimicked the conditions that will exist in Super-K once Gd sulfate is loaded (SK-Gd), this was the definitive test to demonstrate the capabilities of the band-pass system and the null effects on detector components. Water transparency was measured at least daily and Gd sulfate concentration was monitored with variable frequency (more often while conditions were changing and more sparsely  when in stable conditions). Both transparency and concentration are measured from three sampling ports inside EGADS. These are at the bottom, center and top sections of the detector, see Figure~\ref{fig:EGADS-tank-views}. The results are shown in Figure~\ref{fig:Transparency-Concentration}. The blue band represents the typical LL15 values for Super-K's ultrapure water over the last decade; these are between 75$\%$ and 82$\%$, left y-axis scale. EGADS' LL15 values as a function of the sampling date are the upper blue, red, and green lines for the bottom, center, and top sampling ports, respectively. As described in Subsection~\ref{subsec:band-pass}, the water purification system takes water from the top and injects it back into EGADS from the bottom. As a consequence LL15 values are expected to be slightly larger in the bottom than at the top. The lower blue, red, and green lines are the measured Gd sulfate concentrations on a given sampling date for the bottom, center, and top sampling ports, respectively. The concentration scale can be read on the right and the horizontal dashed line indicates the final expected value. The vertical hatched areas indicate events for which the conditions changed in the detector operation; a short description is added.

At the end of November 2014 the first Gd sulfate loading took place in this final version of the detector. The goal was to reach a concentration of 0.02$\%$ Gd sulfate. The LL15 had a sudden drop below the blue band at all sampling positions. After that the LL15 recovered as water was circulated, and the LL15 values returned into the typical Super-K ultrapure range. The Gd sulfate concentration increased in all sampling positions and quickly became homogeneous throughout the detector.

Three additional Gd sulfate loadings followed: end of January 2015 (0.1$\%$ Gd sulfate), middle April 2015 (0.16$\%$ Gd sulfate), and end of April 2015 (0.2$\%$ Gd sulfate). All loadings demonstrated a similar pattern to the first one: sudden LL15 drop followed by a rapid recovery back to the typical Super-K transparency values, while the Gd concentration as measured by the AAS rose homogeneously throughout the detector to the expected values. 

%%%%%%%%%%%%%%%%%%%%%%%%%%%%%%%%%%%%%%%%%%%%%%%%%%%%%%%%%%%%%%%%%%%%%%%%%
%%%%%%%%%%%%%%%%%%%%%%%%%%%%%%%%%%%%%%%%%%%%%%%%%%%%%%%%%%%%%%%%%%%%%%%%%%

A primary result is that the LL15 values stay within the blue band if no disruptive event happens. In time-sequential order these events were: fast recirculation system was turned off for a test to see the impact of this system in LL15 (after that it was turned back on), band-pass system off (to test its impact in LL15), 2016's annual water system maintenance, band-pass and fast recirculation systems off (due to an unexpected system failure), and 2017's water system maintenance. In each case the LL15 values drop but after a short period of stable running the values return to the typical Super-K values. Note that the Gd sulfate concentration remains, within measurement uncertainties,  constant around the final target concentration. This means that we could not detect any Gd loss after two and a half years of running EGADS at full concentration, which involved passing the entire water volume through the water systems 650 times.

\subsection{Main lessons from running EGADS}

In summary, after full Gd-loading, the water transparency is within typical Super-K values, which means that Gd sulfate is essentially transparent to Cherenkov light. The water filtration system can achieve and indefinitely maintain good water quality. In addition, the Gd losses are minimal and we could not detect them after more than two years operating at full concentration. The Gd sulfate quickly dissolves and is homogeneously distributed throughout the detector (no stratification).

Although there was no sign of deterioration of any detector component, EGADS was emptied in November 2017 to perform an eye-check of the inner structure, PMTs, etc. After these two and a half years under conditions designed to be as close as possible to those that will exist at SK-Gd, we found no sign of deterioration in any of the detector components in what constitutes the most realistic possible soak test.

\section{Backgrounds in Gd sulfate}

The number of free neutrons in the ultrapure Super-K water is very low. However, when adding Gd sulfate these neutrons will become visible. Further, impurities in the Gd sulfate may decay. The decay of these impurities create more neutrons as well as alpha, gamma and beta particles that, alone or in coincidence with other events may mimic signals in low energy analyses. Thus, the radiopurity of Gd sulfate has to be taken into account. %and in addition, we may be adding an important neutron source if the Gd sulfate radiopurity is not adequate. 

In Table~\ref{tab:radioactive_contamination} we show the relevant radioactive chains (first and second columns) and the typical activities per kilogram of Gd sulfate powder (third column) from Stanford Materials. We studied these chains and their potential impact on DSNB and solar neutrino analyses by mimicking their signals once Gd sulfate is mixed. The calculated upper limits that would be tolerable for our physics goals are indicated in the fourth and fifth columns for the most restrictive cases for DSNB and solar analyses, respectively. If no number is given (-) it means that the corresponding requirement is less restrictive.

%%%%%%%%%%%%%%%%%%%%%%%%%%%%%%%%%%%%%%%%%%%%%%%%%%%%%%%%%%%%%%%%%%%%%%%%%
\begin{table}[h]
~~All units in mBq/Kg.
\vspace{-0.2cm}
\begin{center}
\renewcommand\arraystretch{1.1}  % vertical   separation
\renewcommand{\tabcolsep}{5pt}   % horizontal separation
%\newcolumntype{C}{>{\centering\arraybackslash}m{1cm} }
\begin{tabular}{c|c|c|cc|c|c|c}  
    Chain   & Part of the           &  Typical & DSNB  & Solar    & Company  &  Company  & Company  \\ 
            &  chain                &          &       &          &    A     &     B     &    C     \\ \hline

 $^{238}$U  & $^{238}$U             &  50      & $<$ 5 &   -      & $<$ 0.04 & $<$ 0.04  & $<$ 0.04 \\
            & $^{226}$Ra            &  5       &  -    & $<$ 0.5  & $<$ 0.2  & $<$ 0.2   & ~1       \\ \hline
 $^{232}$Th & $^{232}$Th            &  10      &  -    & $<$ 0.05 & 0.02     & 0.06      & 0.09     \\ 
            & $^{228}$Th            &  100     &  -    & $<$ 0.05 & $<$ 0.3  & $<$ 0.26  & ~2       \\ \hline
 $^{235}$U  & $^{235}$U             &  32      &  -    & $<$ 30   & $<$ 0.4  & $<$ 0.3   & $<$ 1.3  \\ 
            & $^{227}$Ac/$^{227}$Th &  300     &  -    & $<$ 30   & $<$ 1.5  & $<$ 1.2   & $<$ 3.1  \\ \hline
\hline
\end{tabular}
\caption{Relevant radioactive contamination and typical impurities in untreated Gd sulfate and the requirements from  DSNB and solar neutrino physics goals. Units in this table are in mBq/Kg.}
\label{tab:radioactive_contamination}
\end{center}
\end{table}
%%%%%%%%%%%%%%%%%%%%%%%%%%%%%%%%%%%%%%%%%%%%%%%%%%%%%%%%%%%%%%%%%%%%%%%%%%%

We have been collaborating with several chemical companies to produce a Gd sulfate powder that meets these requirements. They produce highly radiopure Gd sulfate and we test the radiopurity with low background germanium detectors at Canfranc (Spain) and Boulby (UK), as well as with ICP/MS measurements at Kamioka (Japan)~\cite{Ito:2017zzt}. The radioactive contamination levels achieved by these companies are shown in the last three columns of Table~\ref{tab:radioactive_contamination}. There is one company that already meets the requirements within our measurement uncertainties. There are others that are still working on increasing the radiopurity of their Gd sulfate powder.

\section{Gd removal}

To achieve a concentration of 0.2$\%$ Gd sulfate in Super-K we will need to dissolve about 100 tons of Gd sulfate powder. Someday we will need to empty the tank and remove the dissolved Gd sulfate, at least at the end of the SK-Gd phase. We have been investigating methods to remove Gd in a quick, as well as cost effective, manner. These methods range from filter presses and precipitation via pH control to cyclonic separation or spillover filtration among others.

We found that the most effective and straightforward method to deploy is the use of a cation ion-exchange resin. This resin releases three Na$^+$ ions for every captured Gd$^{3+}$ ion. The highly-charged Gd$^{3+}$ ion is then tightly bound to the resin matrix and cannot accidentally escape once it has been captured. The only method to retrieve the captured Gd is to flush the resin with concentrated acid. Used resin is inert and stable, and has been designed for safe transport.

This resin has been successfully tested at EGADS. We analyzed the treated water with an ICP/MS and we could not find any Gd trace. Taking into account the ICP/MS uncertainties, we determined that the Gd concentration has been reduced from one part per thousand in the feedstock to less than 0.5 parts per billion in the waste stream.

\section{Summary}

As proposed by GADZOOKS!, WC detectors would greatly benefit from efficient neutron tagging capabilities. Adding Gd while maintaining good water quality will make this possible in large detectors like Super-K, and EGADS has demonstrated the feasibility of this technique. Gd sulfate is essentially transparent to Cherenkov light and dissolves easily and homogeneously; Gd sulfate can be produced to the required levels of radio-purity needed for our studies; finally, it can be  removed in an efficient and economical way whenever needed. 

The success of EGADS showing the feasibility of this technique was key to the decision made by the Super-Kamiokande collaboration to refurbish the Super-K detector in 2018 as a first step to load Gd sulfate and benefit from effective neutron tagging capabilities.

\Acknowledgements
We  gratefully  acknowledge  the  cooperation  of the  Kamioka  Mining  and  Smelting  Company.
This work was supported by the JSPS KAKENHI Grant Numbers JP21224004, JP26000003, JP24103004 and JP17H06365. Funding support was provided by Kavli IPMU (WPI), the University of Tokyo and the US Department of Energy. We thank the "Consorcio Laboratorio Subterraneo de Canfranc" (Spain) and the Boulby Underground Research Laboratory and in particular the staff of the BUGS facility (UK) for supporting the low-background materials screening work. Some of us have been supported by funds from the European Union H2020 RISE-GA641540-SKPLUS, the Ministry of Education (2018R1D1A3B07050696, 2018R1D1A1B07049158), the Science and Technology Facilities Council (STFC) and GridPP, UK, and the European Union’s H2020-MSCA-RISE-2018 JENNIFER2 grant agreement no.822070.

\bibliography{biblio}
\bibliographystyle{plain}

\end{document}

%% file: econfmacros.tex
\def\beq{\begin{equation}}
\def\eeq#1{\label{#1}\end{equation}}
\def\eeqn{\end{equation}}

\def\beqa{\begin{eqnarray}}
\def\eeqa#1{\label{#1}\end{eqnarray}}
\def\eeqan{\end{eqnarray}}

\let\bar=\overbar

\def\Dslash{\not{\hbox{\kern-4pt $D$}}}
\def\dslash{\not{\hbox{\kern-2pt $\del$}}}

\def\msb{{\bar{\ssstyle M \kern -1pt S}}}

%% file: authorList.tex
\newcommand{\AFFicrr}{\affil[1]{Kamioka Observatory, Institute for Cosmic Ray Research, University of Tokyo, Kamioka, Gifu 506-1205, Japan}}
\newcommand{\AFFipmu}{\affil[2]{Kavli Institute for the Physics and Mathematics of the Universe (WPI), The University of Tokyo Institutes for Advanced Study}}
\newcommand{\AFFuci}{\affil[3]{Department of Physics and Astronomy, University of California, Irvine, Irvine, CA 92697-4575, USA}}
\newcommand{\AFFokayama}{\affil[4]{Department of Physics, Okayama University, Okayama, Okayama 700-8530, Japan}}
\newcommand{\AFFautonomaM}{\affil[5]{Department of Theoretical Physics, University Autonoma Madrid, 28049 Madrid, Spain}}
\newcommand{\AFFLSC}{\affil[6]{Laboratorio Subterraneo de Canfranc, 22880 Canfranc Estacion, Huesca, Spain }}

\newcommand{\AFFimperial}{\affil[7]{Department of Physics, Imperial College London , London, SW7 2AZ, United Kingdom}}
\newcommand{\AFFoxford}{\affil[8]{Department of Physics, Oxford University, Oxford, OX1 3PU, United Kingdom}}
\newcommand{\AFFstfc}{\affil[9]{STFC, Rutherford Appleton Laboratory, Harwell Oxford, and Daresbury Laboratory, Warrington, OX11 0QX, United Kingdom}}
\newcommand{\AFFliverpool}{\affil[10]{Department of Physics, University of Liverpool, Liverpool, L69 7ZE, United Kingdom}}
\newcommand{\AFFkings}{\affil[11]{Department of Physics, King’s College London, London,WC2R 2LS, United Kingdom}}
\newcommand{\AFFsheffield}{\affil[12]{Department of Physics and Astronomy, University of Sheffield, S3 7RH, Sheffield, United Kingdom}}
\newcommand{\AFFep}{\affil[13]{Ecole Polytechnique, IN2P3-CNRS, Laboratoire Leprince-Ringuet, F-91120 Palaiseau, France}}
\newcommand{\AFFkek}{\affil[14]{High Energy Accelerator Research Organization (KEK), Tsukuba, Ibaraki 305-0801, Japan}}
\newcommand{\AFFtokai}{\affil[15]{Department of Physics, Tokai University, Hiratsuka, Kanagawa 259-1292, Japan}}
\newcommand{\AFFkobe}{\affil[16]{Department of Physics, Kobe University, Kobe, Hyogo 657-8501, Japan}}
\newcommand{\AFFkicrr}{\affil[17]{Research Center for Cosmic Neutrinos, Institute for Cosmic Ray Research, University of Tokyo, Kashiwa, Chiba 277-8582, Japan}}
\newcommand{\AFFkyoto}{\affil[18]{Department of Physics, Kyoto University, Kyoto, Kyoto 606-8502, Japan}}
\newcommand{\AFFnagoya}{\affil[19]{Institute for Space-Earth Enviromental Research, Nagoya University, Nagoya, Aichi 464-8602, Japan}}
\newcommand{\AFFtokyo}{\affil[20]{Department of Physics, University of Tokyo, Bunkyo, Tokyo 113-0033, Japan}}
\newcommand{\AFFtsingua}{\affil[21]{Department of Engineering Physics, Tsinghua University, Beijing, 100084, China}}
\author[1]{Ll.~Marti}
\author[1]{M.~Ikeda}
\author[1]{Y.~Kato}
\author[1,2]{Y.~Kishimoto}
\author[1,2]{M.~Nakahata}
\author[1,2]{Y.~Nakajima}
\author[1]{Y.Nakano}
\author[1,2]{S.~Nakayama}
\author[1]{Y.~Okajima}
\author[1]{A.~Orii}
\author[1]{G.~Pronost}
\author[1,2]{H.~Sekiya}
\author[1,2]{M.~Shiozawa}
\author[1]{H.~Tanaka}
\author[1]{K.~Ueno}
\author[1]{S.~Yamada}
\author[1]{T.~Yano}
\author[1]{T.Yokozawa}
\AFFicrr

\author[2]{M.~Murdoch}
\author[2]{J.~Schuemann}
\author[2,3]{M.~R.~Vagins}
\AFFipmu

\author[3]{K.~Bays}
\author[3]{G.~Carminati}
\author[3]{N.~J.~Griskevich}
\author[3]{W.~R.~Kropp}
\author[3]{S.~Locke}
\author[3]{A.~Renshaw}
\author[3,2]{M.~B.~Smy}
\author[3]{P.~Weatherly}
\AFFuci

\author[4]{S.~Ito}
\author[4]{H.~Ishino}
\author[4]{A.~Kibayashi}
\author[4,2]{Y.~Koshio}
\author[4]{T.~Mori}
\author[4]{M.~Sakuda}
\author[4]{R.~Yamaguchi}
\AFFokayama

\author[5]{P.~Fernandez}
\author[5]{L.~Labarga}
\AFFautonomaM

\author[6]{I.~Bandac}
\author[6]{J.~Perez}
\AFFLSC

\author[7]{J.~Amey}
\author[7]{R.~P.~Litchfield}
\author[7]{A.~Sztuc}
\author[7]{Y.~Uchida}
\author[7]{W.~Y.~Ma}
\AFFimperial

\author[8,2]{A.~Goldsack}
\author[8,2]{C.~Simpson}
\author[8,9]{D.~Wark}
\AFFoxford
\AFFstfc

\author[10]{L.~H.~V.~Anthony}
\author[10]{N.~McCauley}
\author[10]{A.~Pritchard}
\AFFliverpool

\author[11]{F.~Di~Lodovico}
\author[11]{B.~Richards}
\AFFkings

\author[12]{A.~Cole}
\author[12]{M.~Thiesse}
\author[12]{L.~Thompson}
\AFFsheffield

\author[13]{J.~Imber}
\AFFep

\author[14]{S.~V.~Cao}
\AFFkek

\author[15]{K.~Ito}
\AFFtokai

\author[16,2]{Y.~Takeuchi}
\AFFkobe

\author[17]{R.~Akutsu}
\author[17]{Y.~Nishimura}
\author[17]{K.~Okumura}
\AFFkicrr

\author[18]{S.~Hirota}
\AFFkyoto

\author[19]{F.~Muto}
\AFFnagoya

\author[20,2]{M.~Yokoyama}
\author[20]{Y.~Suda}
\AFFtokyo

\author[21]{H.~Zhang}
\AFFtsingua